\documentclass[12pt]{article}

\usepackage{amssymb,amsmath}

\usepackage{graphicx}
\DeclareGraphicsExtensions{.pdf,.png,.jpg}

\renewcommand{\theequation}{\arabic{section}.\arabic{equation}}
 \catcode`\@=11 \@addtoreset{equation}{section}\catcode`\@=12
\newcommand{\R}{ {\mathbb R} }

\newcommand{\fnm}{\footnotemark}
\newcommand{\fnt}{\footnotetext}

\begin{document}

 \begin{center}

 \large \bf 
On stable  exponential cosmological solutions with 
 two factor spaces in $(1+ m + 2)$-dimensional EGB model
 with $\Lambda$-term
  
  \end{center}

 \vspace{0.3truecm}

\begin{center}

 V. D. Ivashchuk$^{1,2}$ and A. A. Kobtsev$^{3}$ 

\vspace{0.3truecm}

 \it $^{1}$ Institute of Gravitation and Cosmology,
   Peoples' Friendship University of Russia (RUDN University),
   6 Miklukho-Maklaya St.,  Moscow 117198, Russian Federation \\ 

 \it $^{2}$ Center for Gravitation and Fundamental Metrology,
 VNIIMS, 46 Ozyornaya St., Moscow 119361,  Russian Federation
 
 \it $^{3}$ Moscow, Troitsk, 142190,  Russian Federation Institute 
  for Nuclear Research of the Russian Academy of Sciences

\end{center}

\begin{abstract}

A  $(m+ 3)$-dimensional Einstein-Gauss-Bonnet  gravitational model including  the Gauss-Bonnet term and the cosmological term $\Lambda$ is considered.  Exact solutions with  exponential time dependence of two scale factors, governed by two Hubble-like parameters $H >0$ and $h \neq H$, corresponding to factor spaces of dimensions $m >2$ and $l = 2$, respectively, are found. 
Under certain restrictions on $x = h/H $,  the stability of the solutions  in a class of cosmological solutions with diagonal  metrics is  proved. 
 A subclass of solutions with small enough variation of the effective gravitational constant $G$ is considered  and the stability of all solutions from this subclass is shown.

\end{abstract}


\newpage

\section{Introduction}

This paper  is a continuation of our previous work \cite{IvKob-GRG-18} devoted to cosmological solutions  with two factor-spaces in the   Einstein-Gauss-Bonnet (EGB) gravitational model 
in $D$ dimensions, which contains  the so-called Gauss-Bonnet term  and cosmological term $\Lambda$.
\fnm \fnt{For more general ``cosmological-type'' case see also ref. \cite{Iv-GC-20}.}  
Such model may be also referred  as EGB$\Lambda$ one.
The Gauss-Bonnet term appeared in low-energy limit  of certain effective string action  \cite{Zwiebach}-\cite{GW}. It also  appears in the  action of Einstein-Lovelock (EL) gravity \cite{L,Lov}
for $D > 4$ as a first non-trivial addition (quadratic in curvature) to the Einstein-Hilbert  action (with a $\Lambda$-term).

Currently, there exists  a certain interest 
to EGB and EL gravitational models and  its extensions, see   \cite{Ishihara}-\cite{Pavl-p-18} and refs. therein.
The main motivation for this activity  is in possible explanation of supernovae (type Ia) observational data \cite{Riess,Perl,Kowalski}, which tell us about  accelerating  expansion of the Universe.
There exists also  a considerable interest in studying of black-hole solutions in EGB and  EL 
theories, see refs. \cite{BD}-\cite{GG}.

 Here we study the cosmological solutions  with exponential dependence of scale factors upon synchronous time variable and find  
a new class of  solutions with  two scale factors, governed by two Hubble-like parameters $H >0$ and $h$, which correspond to factor spaces of dimensions $m > 2$ and $2$, respectively, ($D = 1 + m + 2$)
and obey relations: $m H + 2 h \neq 0$ and  $H \neq h$. 
Any of  these solutions  describes   an exponential  expansion  of $3$-dimensional subspace with Hubble parameter $H > 0$ \cite{Ade}.  Here, as in our previous paper \cite{IvKob-GRG-18}, we use the Chirkov-Pavluchenko-Toporensky trick of reducing  the set of polynomial equations  on parameters $H$ and $h$ from ref. \cite{ChPavTop1}.  
By using results of  refs. \cite{Ivas-16}  we 
show  that the solutions are stable when certain restrictions on ratio $h/H$ are imposed. 
Here    the stability of the  solutions is understood such that the allowed perturbations of the metric do not  output the solutions from the class of cosmological solutions with diagonal metrics. 
 
 We also study solutions with a small enough  variation  of the effective gravitational constant 
$G$  in Jordan  frame \cite{RZ-98,I-96}
which obey the  restrictions on   $G$-dot from ref. \cite{Pitjeva}. We show that these
solutions are  stable.

Earlier in Ref. \cite{IvKob-GRG-18}  we were dealing with exponential cosmological solutions    in the EGB model (with a $ \Lambda$-term) with two non-coinciding Hubble-like parameters   $H > 0$ and $h$ obeying $S_1 = m H + l h \neq 0$ and corresponding to    $m$- and $l$-dimensional factor spaces with $m > 2$ and $l>2$, respectively. In this case we have found two sets of solutions obeying: a) $\alpha > 0$, $ \Lambda < \alpha^{-1} \lambda_{+}(m,l)$ and  b) $\alpha < 0$, $  \Lambda > |\alpha|^{-1}  \lambda_{-}(m,l)$, with $\lambda_{\pm}(m,l) > 0$. 
 Here $\alpha = \alpha_2 / \alpha_1 $, where $\alpha_1$ and $\alpha_2$ are two non-zero constants of the model.  
  We show,  that in the case $m > 2$, $l = 2$, $\alpha > 0$ the  ``spectrum'' of allowed values for $\Lambda$ is unbounded from the top and the bottom and the ratio $x = h/H$ is unbounded from the bottom ($- \infty < x < x_{*}$), while in the case  $m > 2$, $l > 2$, $\alpha > 0$ \cite{IvKob-GRG-18}  the set  of allowed values for $\Lambda$ is bounded from the top and set of ratios $x = h/H$ is bounded ($x_{-} < x < x_{+} < 0$). From  matematical point of view we have here a more simple task since for  $m > 2$, $l= 2$ the polynomial master equation for $x$ is of third order while in the case $m > 2$, $l > 2$  it is  generically of fourth order \cite{IvKob-GRG-18}. The solution to the master equation is presented in Appendix.

\section{The cosmological model}

We start with the action
\begin{equation}
  S =  \int_{M} d^{D}z \sqrt{|g|} \{ \alpha_1 (R[g] - 2 \Lambda) +
              \alpha_2 {\cal L}_2[g] \},
 \label{r1}
\end{equation}
where $g = g_{MN} dz^{M} \otimes dz^{N}$ is the metric  on
the manifold $M$, ${\dim M} = D$, $|g| = |\det (g_{MN})|$, $\Lambda$ is
the cosmological term, $R[g]$ is scalar curvature,
$${\cal L}_2[g] = R_{MNPQ} R^{MNPQ} - 4 R_{MN} R^{MN} +R^2$$
is the  Gauss-Bonnet term and  $\alpha_1$, $\alpha_2$ are
nonzero constants of the model.

Here we deal with the manifold
\begin{equation}
   M = \R  \times   M_1 \times \ldots \times M_n 
   \label{r2.1}
\end{equation}
and the metric
\begin{equation}
   g= - d t \otimes d t  +
      \sum_{i=1}^{n} B_i e^{2v^i t} dy^i \otimes dy^i,
  \label{r2.2}
\end{equation}
where   $B_i > 0$ are constants, $i = 1, \dots, n$. In (\ref{r2.1}) 
$M_1, \dots,  M_n$  are one-dimensional manifolds, either  $\R$ or $S^1$, 
and $n > 3$.

The equations of motion for the action (\ref{r1}) and the metric  (\ref{r2.2}) read
 \cite{ErIvKob-16,IvKob-18mm}
\begin{eqnarray}
  G_{ij} v^i v^j + 2 \Lambda
  - \alpha   G_{ijkl} v^i v^j v^k v^l = 0,  \label{r2.3} \\
    \left[ 2   G_{ij} v^j
    - \frac{4}{3} \alpha  G_{ijkl}  v^j v^k v^l \right] \sum_{k =1}^n v^k 
    - \frac{2}{3}   G_{sj} v^s v^j  +  \frac{8}{3} \Lambda = 0, 
                                                           \label{r2.4}
\end{eqnarray}
$i = 1,\ldots, n$, where  $\alpha = \alpha_2/\alpha_1$. 
Here as in refs. \cite{Iv-09,Iv-10} we use the notations
\begin{equation}
G_{ij} = \delta_{ij} -1, 
\qquad   G_{ijkl}  = G_{ij} G_{ik} G_{il} G_{jk} G_{jl} G_{kl}.
\label{r2.4G}
\end{equation}
For  $n > 3$  ($D > 4$) we get a set of forth-order polynomial  equations.

In ref. \cite{ChPavTop} an isotropic solution $v^1 = \ldots = v^n = H$  
 was obtained for $\Lambda \neq 0$,  $\alpha  < 0$, see also ref. \cite{Ivas-16}.

It was proved  in ref. \cite{Ivas-16} that there are no more than
three different  numbers among  $v^1,\dots ,v^n$ obeying $\sum_{i = 1}^{n} v^i \neq 0$. 

\section{Solutions with two Hubble-like parameters}

Here we deal with solutions to  equations (\ref{r2.3}), 
(\ref{r2.4}) governed by two  parameters:
\begin{equation}
  \label{r3.1}
   v =(\underbrace{H,H,H}_{``our" \ space},\underbrace{\overbrace{H, \ldots, H}^{m-3}, 
   h, h}_{internal \ space}).
\end{equation}
The Hubble-like parameter $H$  corresponds  
to  $m$-dimensional factor space with $m > 2$ and  Hubble-like parameter $h$ 
corresponds to  $2$-dimensional factor space. 

Keeping in mind possible cosmological applications, we split the $m$-dimensional  
factor space into the  product of two subspaces of dimensions $3$ and $m-3$, respectively,
and obtain  $3$--dimensional ``our" space and   
$((m - 3) + 2)$-dimensional anisotropic  ``internal space''. For physical applications (in our epoch) the internal space should be chosen to be compact one, i.e. one  should put in (\ref{r2.1}) $M_4 = \ldots = M_n = S^1$, while 
$M_1 = M_2 = M_3 = \R$ and  the internal scale factors corresponding to present time  should be small enough in comparison with the scale factors of ``our'' space.  
 
In order to describe possible accelerated expansion of a
$3d$ subspace (``our Universe'') we put 
\begin{equation}
  \label{r3.2a}
   H > 0. 
\end{equation}
 

According to  ansatz (\ref{r3.1}),  the $m$-dimensional factor space is expanding with the Hubble parameter $H >0$, while  the evolution of the $2$-dimensional factor space is driven by the Hubble-like  parameter $h$.

 As in our previous paper \cite{IvKob-GRG-18}  we impose the following restrictions on parameters $H$ and $h$  
  \begin{equation}
   m H + 2 h \neq 0, \qquad  H \neq h.
   \label{r3.3}
   \end{equation}
  
  With the ansatz  (\ref{r3.1}) and  the restrictions (\ref{r3.3}) imposed
   the relations (\ref{r2.3}) and (\ref{r2.4}) can be reduced to  
  a  set of two polynomial equations \cite{ChPavTop1,Ivas-16} 
  \begin{eqnarray}
  E =  m H^2 + 2 h^2 - (mH + 2h)^2  + 2 \Lambda 
           \nonumber \\
        - \alpha [m (m-1) (m-2) (m - 3) H^4
          \nonumber \\
       + 8 m (m-1) (m-2) H^3 h   
       + 12 m (m-1)  H^2 h^2] = 0,
                \label{r3.4}   \\
  Q =  (m - 1)(m - 2)H^2 + 2 (m - 1) H h  = - \frac{1}{2 \alpha}. \qquad
     \label{r3.5}
  \end{eqnarray}

For $m > 2$  equation  (\ref{r3.5}) implies   
\begin{equation}
H   =     (- 2 \alpha {\cal P})^{-1/2}, 
 \label{r3.6}
\end{equation}
where 
\begin{eqnarray}
{\cal P}  =  {\cal P}(x,m)  =   (m - 1)(m - 2)  + 2 (m - 1) x, 
 \label{r3.7}  \\
    x  = h/H,
    \label{r3.7x}
 \end{eqnarray}
and 
 \begin{equation}
\alpha {\cal P} < 0. 
 \label{r3.8} 
\end{equation}

Restrictions (\ref{r3.3}) may be written as 
\begin{equation}
  x \neq x_d = x_d(m) \equiv  - m/2, \qquad  x \neq x_a \equiv 1.
 \end{equation}
 
 Relation  (\ref{r3.5}) is valid only if        
\begin{equation}
  {\cal P}(x,m) \neq 0. 
 \label{r3.8b}
\end{equation}

Substituting  (\ref{r3.6}) into (\ref{r3.4})
we obtain
\begin{eqnarray}
    \Lambda \alpha = \lambda  =  \lambda(x,m) \equiv 
     \frac{1}{4} ({\cal P}(x,m))^{-1} {\cal M}(x,m)
     \qquad \nonumber \\
      + \frac{1}{8 }( {\cal P}(x,m))^{-2} {\cal R}(x,m),  
              \qquad    \label{r3.8L}  \\
   {\cal M}(x,m) \equiv  m  + 2 x^2 -(m  + 2 x)^2, 
              \qquad   \label{r3.8M}  \\
   {\cal R}(x,m) \equiv  m (m-1) (m-2) (m - 3)                 
                \qquad  \nonumber \\    
     + 8 m (m-1) (m-2) x   + 12 m (m-1) x^2 . 
                \qquad   \label{r3.8R}  
\end{eqnarray}

Using (\ref{r3.8b}) we get
 \begin{eqnarray} 
  x \neq x_{*} = x_{*}(m)   \equiv - (m - 2)/2.  
            \qquad \label{r3.9D}
 \end{eqnarray}

We have ($m >2$)
 \begin{equation} 
     x_{*}(m) < 0.   \label{r3.11} 
  \end{equation}

Using (\ref{r3.8}) and (\ref{r3.8L}) we find 
 \begin{equation}
   \Lambda   = \alpha^{-1} \lambda(x,m),   
               \qquad \label{r3.13a} 
 \end{equation}
where 
\begin{equation} 
    x <  x_{*}(m) \ {\rm for} \  \alpha > 0  
               \qquad \label{r3.13b}
 \end{equation}    
  and    
   \begin{equation}  
     x > x_{*}(m) \ {\rm for} \  \alpha < 0.
                 \qquad \label{r3.13c}
 \end{equation}

For $ \alpha < 0$ we have the following limit 
  \begin{equation}  
  \lim_{x \to + \infty} \lambda(x,m) = - \infty,
                  \qquad \label{r3.13.1}
  \end{equation}
  while for $ \alpha > 0$ we obtain 
    \begin{equation}  
    \lim_{x \to - \infty} \lambda(x,m) = + \infty.
                    \qquad \label{r3.13.2}
    \end{equation}
  
  We note that 
  \begin{equation} 
  \lambda(x) \sim - x/(4(m-1))
  \label{r3.13.a}   
   \end{equation}
   as $x \to \pm \infty$.
  
 For $x = 0$ we get  (here $ \alpha < 0$, $m > 2$)
\begin{equation}  
   \Lambda = \Lambda_{0} = \alpha^{-1} \lambda(0,m) = 
    -  \frac{m(m + 1)}{8 \alpha (m - 1)(m - 2)} > 0,
                  \qquad \label{r3.13.L}
  \end{equation}
 and 
\begin{equation}
  H  = H_0 = (- 2 \alpha (m - 1)(m - 2))^{-1/2}, \qquad h = 0,  
 \label{r3.13.H.0}
 \end{equation}  
i.e. (see (\ref{r2.1}),  (\ref{r2.2})) we get the product
of (a part of) $(m+1)$-dimensional de-Sitter space and $2$-dimensional 
Euclidean space.

{\bf ``Master'' equation.} Rewriting eq. (\ref{r3.8L})  
we are led to a ``master'' equation
\begin{equation}
       2 {\cal P}(x,m) {\cal M}(x,m)
       +  {\cal R}(x,m) - 8 \lambda ( {\cal P}(x,m))^{2} = 0,  
              \qquad    \label{r3.13.M} 
\end{equation}                
which is of third order in $x$. 
For any $m > 2$  the  equation (\ref{r3.13.M}) may be readily solved  in radicals,
the solution is presented in Appendix A. 

The behaviour of the function  $\lambda(x,m)$   in the vicinity of 
 the point  $x_{*}(m)$ is desribed by the following proposition.

{\bf Proposition 1.} {\em For $m > 2$,  
 \begin{equation}
    \lambda(x,m) \sim B (m) (x - x_{*}(m))^{-2}, 
 \label{r3.13.H}
\end{equation}
 as $x \to x_{*} = x_{*}(m)$, where $B(m) = - (m(m-2))/(32(m - 1))  < 0$ and hence
 \begin{equation}
   \lim_{x \to x_{*}}  \lambda(x,m) = - \infty. 
  \label{r3.13.lim}
 \end{equation}
 }

For a given $m > 2$ let us analyze the behaviour of the function  $\lambda(x,m)$ in $x$ for 
 $x \neq x_{*}(m)$.
First, we find the  extremum  points  obeying $\frac{\partial}{\partial x} \lambda(x,m) = 0$.
We obtain 
\begin{eqnarray} 
  \frac{\partial}{\partial x} \lambda(x,m) = - f(x,m) ({\cal P}(x,m))^{-3},
                \qquad \label{r3.14a} \\
        f(x,m) = (m-1)^2 (x-1)(2x+m)(x + m-2),
                 \qquad \label{r3.14b}
\end{eqnarray}
$x \neq x_{*}(m)$. Using these relations we get the following
extremum points
 \begin{eqnarray} 
         x_a = 1,
            \qquad \label{r3.15a} \\
        x_c = x_c(m) \equiv - (m-2) < 0,
                     \qquad \label{r3.15c} \\
        x_d = x_d(m) \equiv - \frac{m}{2} < 0.
                             \qquad \label{r3.15d}       
 \end{eqnarray}

   For  points $x_c, x_d$  the following  inequalities are valid
  \begin{equation} 
                  x_i(m) < x_{*}(m),    \qquad \label{r3.17x}
  \end{equation}
  $i = c,d$  ($m >2$). 

We also get 
\begin{eqnarray}
(1) \  x_c < x_d & \text{for }  m > 4,  \label{r3.20a} \\
(2) \  x_c = x_d & \text{for }  m = 4, \label{r3.21a}   \\
(3) \  x_d < x_c & \text{for }  m = 3 \label{r3.20b}.  \\
\end{eqnarray}

For $\lambda_i=\lambda(x_i,m)$, $i=a,c,d$, we obtain
\begin{equation}
\lambda_a = -\frac{(m + 1)(m+2)}{8(m-1)m}<0, \label{r3.22a}
\end{equation}

\begin{equation}
\lambda_c = \frac{3 m^2 - 13 m + 16}{8(m-2)(m-1)}>0, \label{r3.22c}
\end{equation}
and 
\begin{equation}
\lambda_d= \frac{m(m+2)}{32(m - 1)}>0 \label{r3.22d}
\end{equation}
for all $m > 2$.

We also get 
\begin{equation}
\lambda_d - \lambda_c=\frac{(m - 4)^3}{32(m-2)(m -1)} 
             \label{r3.25a} 
\end{equation}
  for  $m > 2$. By using this relation we obtain 
\begin{equation}
\lambda_d - \lambda_c \begin{cases}
 >0, \ \text{if } m > 4, \\
 =0, \ \text{if } m = 4, \\
 <0, \ \text{if } m = 3.
\end{cases}   \label{r3.26}
\end{equation}

Let us  denote by $n(\Lambda, \alpha)$ the number of solutions 
(in $x$) of the equation  $\Lambda \alpha = \lambda(x,m)$. 
We calculate $n(\Lambda, \alpha)$
by using unequalities for points of extremum  $x_i$ 
and $\lambda_i$ ($i = c,d$) presented above and 
relations (\ref{r3.13.1}), (\ref{r3.13.2}), (\ref{r3.13.lim}), (\ref{r3.14a}) and  (\ref{r3.14b}).	

First, we start with the case $\alpha > 0$ and $x < x_{*}$. 
 
{\bf (1) $m > 4$}. We get $ x_c < x_d$ and  $\lambda_c < \lambda_d$.
Here  $x_d$ is point of local maximum 
and   $x_c$ is a point of local minimum.  We obtain          
 \begin{equation}
   n(\Lambda, \alpha) = \begin{cases}
   1, \   \Lambda \alpha \geq  \lambda_d, \\
   3, \  \lambda_c < \Lambda \alpha < \lambda_d, \\
   2, \   \Lambda \alpha = \lambda_c, \\
   1, \   \Lambda \alpha < \lambda_c.
 \end{cases}   \label{r3.28}
 \end{equation}
Here and in what follows we use $x \neq x_d$.

We present two functions $y = \lambda(x)$ (with $\alpha > 0$ and $\alpha < 0$) for   $m= 5$   at
Figure 1. In this case  we have  $x_{*} = -3/2$, $x_c = -3$, $x_d = - 5/2$, $\lambda_c = 13/48 \approx 0.2708$ and 
$ \lambda_d = 35/128 \approx 0.2734$ and $ \lambda_a = - 21/80 = - 0.2625$. 
At this and other figures  
the  left branch
corresponds to $\alpha > 0$ ($x < x_{*}$) and the right branch  corresponds to $\alpha < 0$ ($x > x_{*}$).

\begin{figure}[!h]
	\begin{center}
		\includegraphics[width=0.75\linewidth]{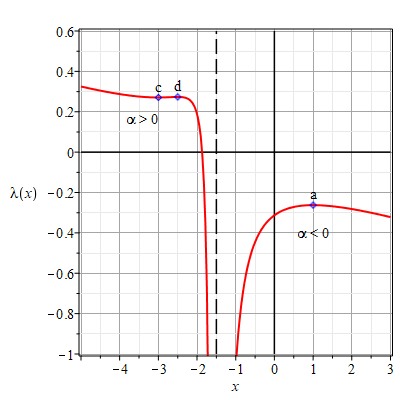}
		\caption{The functions $y = \lambda(x)$ for $\alpha > 0$ (left) and  $\alpha < 0$
		(right) and   $m= 5$.}
\label{rfig:1}
	\end{center}
\end{figure}

 {\bf (2) $m = 4$}. We have  $x_{*} = -1$, $x_c = x_d = -2$, $\lambda_c = \lambda_d = 1/4$ 
 and $\lambda_a = - 5/16 = - 0.3125$. Here   $x_c = x_d = -2$  is the point of inflection.
     We obtain          
  \begin{equation}
    n(\Lambda, \alpha) = \begin{cases}
    1, \  \Lambda \alpha > \lambda_d, \\
     0, \   \Lambda \alpha = \lambda_d, \\
      1, \   \Lambda \alpha < \lambda_d. 
   \end{cases}   \label{r3.29}
  \end{equation}

The functions $y = \lambda(x)$ for   $m= 4$ are depicted at Figure 2.

\begin{figure}[!h]
	\begin{center}
		\includegraphics[width=0.75\linewidth]{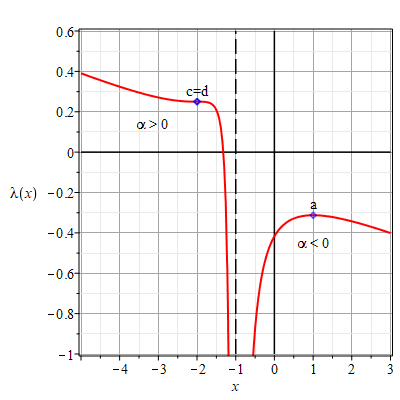}
		\caption{The functions $y = \lambda(x)$ for $\alpha > 0$ (left) and  $\alpha < 0$
				(right) and  $m= 4 $.}
		\label{rfig:2}
	\end{center}
\end{figure}

\pagebreak

 {\bf (3) $ m = 3$.}  The functions $y = \lambda(x)$ for  $m= 3$ are depicted at
 Figure 3.  We have  $x_{*} = -1/2$, $x_c = -1$, $x_d = - 3/2$, $\lambda_c = 1/4$,  
 $\lambda_d = 15/64 \approx 0.2344$ and $\lambda_a = - 10/24$.
 In this subcase we have $\lambda_d < \lambda_c$ and hence 
 
  \begin{equation}
    n(\Lambda, \alpha) = \begin{cases}
    1, \  \Lambda \alpha > \lambda_c, \\
    2, \   \Lambda \alpha = \lambda_c, \\
    3, \  \lambda_d < \Lambda \alpha < \lambda_c, \\
    1, \   \Lambda \alpha  \leq  \lambda_d. \\
   
  \end{cases}   \label{r3.30}
  \end{equation}

 \begin{figure}[!h]
 	\begin{center}
 		\includegraphics[width=0.75\linewidth]{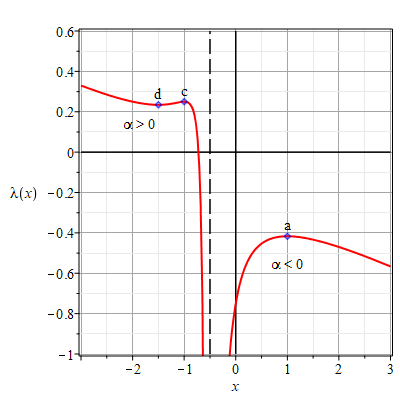}
 		\caption{The functions $y = \lambda(x)$ for $\alpha > 0$ 
 		 (left) and  $\alpha < 0$ (right) and  $m= 3$.}
 		\label{rfig:3}
 	\end{center}
 \end{figure}

{\bf Bounds on  $\Lambda \alpha$ for $\alpha > 0.$}
If we summarize all cases presented above we get that for $\alpha > 0$
exact solutions under consideration exist for all $\Lambda$
 \begin{equation}
    - \infty < \Lambda \alpha < + \infty, 
      \label{r3.34s}
   \end{equation}
when $m \neq 4$, while the extra  restriction
 \begin{equation}
    \qquad \Lambda \alpha \neq \lambda_d.
      \label{r3.34sd}
 \end{equation}
should be imposed for $m =4$.
 
Let us consider the case  $\alpha < 0$. 
We have $\Lambda \alpha =  \lambda(x) $, where  $x > x_{*}$.  Due to  
the relations (\ref{r3.13.1}), (\ref{r3.13.lim}), (\ref{r3.14a}) and  (\ref{r3.14b})
the function $ \lambda(x)$  is monotonically increasing in the  interval  
$(x_{*}, x_{a}= 1)$ from $- \infty$ to $ \lambda_{a} $ and
 it is monotonically decreasing in the interval 
$(1, +\infty)$ from $ \lambda_{a} $ to $- \infty$. 

Here $x_{a}$ is a point of local maximum of the function $\lambda(x) $. This
point is excluded from the solution.  
 
 The  functions  $y = \lambda(x)$   for $\alpha < 0$  and  $m = 5, 4, 3$  
are presented at Figures 1, 2, 3, respectively (at right panels: $x >  x_{*}(m)$).

   For the number of solutions (for $\alpha < 0$) we obtain          
   \begin{equation}
     n(\Lambda, \alpha) = \begin{cases}
     2, \  \Lambda (-\alpha) > |\lambda_{a}|, \\
     0, \   \Lambda (-\alpha) \leq |\lambda_{a}|. \\
    \end{cases}   \label{r3.36}
   \end{equation}
Here  $x \neq x_a =1$. Hence, for $\alpha < 0$ and big enough values of $\Lambda$
there exists  two solutions $x_1, x_2$:  $x_{*} < x_1 <  x_2 $.

{\bf Bounds on  $\Lambda (-\alpha)$ for $\alpha < 0.$}
It follows from (\ref{r3.36}) that for $\alpha < 0$
exact solutions  under consideration exist if and only if
 \begin{equation}
    \Lambda (-\alpha) >     |\lambda_a| = \frac{(D -2)(D-1)}{8(D - 4)(D - 3)}, 
              \label{r3.36s}
   \end{equation}
  see  (\ref{r3.22a}). This relation is valid for all  $D = m + 3 > 5$ ($m > 2$).

\section{Stability analysis}

Here, 
we study the stability of exponential solutions (\ref{r2.2}).
In what follows we use the results of refs. \cite{Ivas-16},
when the total volume factor is non-static, i.e. if
\begin{equation}
  S_1(v) = \sum_{i = 1}^{n} v^i \neq 0.
  \label{r4.1}
\end{equation}

As in  \cite{Ivas-16}, we  impose the following restriction 
\begin{equation}
  ({\rm R }) \quad  \det (L_{ij}(v)) \neq 0
  \label{r4.2}
\end{equation}
where  
\begin{equation}
L =(L_{ij}(v)) = (2 G_{ij} - 4 \alpha G_{ijks} v^k v^s).
   \label{r4.1b}
 \end{equation}

Here we remind that the cosmological ansatz with the (diagonal) metric 
\begin{equation}
 g= - dt \otimes dt + \sum_{i=1}^{n} e^{2\beta^i(t)}  dy^i \otimes dy^i,
 \label{r4.3}
\end{equation}
gives us  the  set of  equations  \cite{ErIvKob-16,Ivas-16,IvKob-18mm}
\begin{eqnarray}
     E = G_{ij} h^i h^j + 2 \Lambda  - \alpha G_{ijkl} h^i h^j h^k h^l = 0,
         \label{r4.3.1} \\
         Y_i =  \frac{d L_i}{dt}  +  (\sum_{j=1}^n h^j) L_i -
                 \frac{2}{3} (G_{sj} h^s h^j -  4 \Lambda) = 0
                     \label{r4.3.2a}
          \end{eqnarray}
(for $\Lambda = 0$ see \cite{Iv-09,Iv-10}), where $h^i = \dot{\beta}^i$,  and         
 \begin{equation}
  L_i = L_i(h) = 2  G_{ij} h^j
       - \frac{4}{3} \alpha  G_{ijkl}  h^j h^k h^l  
       \label{r4.3.3},
 \end{equation}
 $i = 1,\ldots, n$.

It was proved in ref.  \cite{Ivas-16} that a constant solution
$(h^i(t)) = (v^i)$ ($i = 1, \dots, n$; $n >3$) to eqs. (\ref{r4.3.1}), (\ref{r4.3.2a})
obeying restrictions (\ref{r4.1}), (\ref{r4.2}) is  stable under perturbations
\begin{equation}
 h^i(t) = v^i +  \delta h^i(t), 
\label{r4.3h}
\end{equation}
 $i = 1,\ldots, n$,  as $t \to + \infty$  if
\begin{equation}
   S_1(v) = \sum_{k = 1}^{n} v^k > 0
 \label{r4.1a}
\end{equation}
and  it is unstable as $t \to + \infty$ if 
\begin{equation}
   S_1(v) = \sum_{k = 1}^{n} v^k < 0.
 \label{r4.1c}
\end{equation}

In our case $ S_1(v) = m H + 2 h$,  hence, due to $H >0$  the restriction
(\ref{r4.1a}) may be written as
\begin{equation}
  x > - \frac{m}{2} = x_d,
 \label{r4.1x}
\end{equation}
while the  restriction (\ref{r4.1c}) may be presented as
\begin{equation}
  x < - \frac{m}{2} = x_d.
 \label{r4.1y}
\end{equation}

In the linear approximation the  perturbations $\delta h^i$ 
obey  the following set of linear equations \cite{Ivas-16}
 \begin{eqnarray}
   C_i(v) \delta h^i = 0, \label{r4.2C} \\
   L_{ij}(v) \delta \dot{h}^j =  B_{ij}(v) \delta h^j.
  \label{r4.3LB}
 \end{eqnarray}
  Here
 \begin{eqnarray}
 C_i(v)  =  2 v_{i} - 4 \alpha G_{ijks}  v^j v^k v^s, \label{r4.3.4} \\
 L_{ij}(v) = 2 G_{ij} - 4 \alpha G_{ijks} v^k v^s,
    \label{r4.3.5} \\
 B_{ij}(v) = - (\sum_{k = 1}^n v^k)  L_{ij}(v) - L_i(v) + \frac{4}{3} v_{j},
                     \label{r4.3.6}
 \end{eqnarray}
 where $v_i = G_{ij} v^j$,  $L_i(v) =  2 v_{i} - \frac{4}{3} \alpha  G_{ijks}  v^j v^k v^s$
 and $i,j,k,s = 1, \dots, n$.

For the restrictions (\ref{r4.1}), (\ref{r4.2}) imposed,  the set of  equations 
 (\ref{r4.2C}), (\ref{r4.3LB})  has the following solution \cite{Ivas-16}
   \begin{eqnarray}
       \delta h^i = A^i \exp( -  S_1(v) t ),
       \label{r4.16}  \\
         \sum_{i =1}^{n} C_i(v)  A^i =0,
         \label{r4.16A}
   \end{eqnarray}
    ($A^i$ are constants) $i = 1, \dots, n$.

 For  the vector $v$ from  (\ref{r3.1}), obeying
 relations (\ref{r3.3}), the matrix $L$ has a block-diagonal form  \cite{Ivas-16} 
\begin{equation}
 (L_{ij}) = {\rm diag}(L_{\mu \nu}, L_{\alpha \beta} ),
 \label{r4.5}
\end{equation}
where
\begin{equation}
  (L_{\mu \nu}) =  G_{\mu \nu} (2 + 4 \alpha S_{HH}),
 \label{r4.6HH}   
 \end{equation}
 is $m \times m$ matrix and  
 \begin{equation}
  L_{\alpha \beta} = G_{\alpha \beta} (2 + 4 \alpha S_{hh}). 
  \label{r4.6hh}
\end{equation}
is $2 \times 2$ matrix and
\begin{eqnarray}
  S_{HH} =  (m-2)(m -3) H^2  + 4(m-2)Hh + 2 h^2,
  \label{r4.7}   \\
  S_{hh} = m(m-1)H^2.
  \label{r4.8}
\end{eqnarray}

The matrix (\ref{r4.5}) is invertible  only if  
\begin{eqnarray}
 S_{HH} \neq - \frac{1}{2 \alpha}, \label{r4.9a}
  \\
 S_{hh} \neq - \frac{1}{2 \alpha}.
 \label{r4.9b}
\end{eqnarray}
We remind that 
the $k \times k$ matrix  $(G_{i j}) = (\delta_{i j} -1 )$ is invertible for $k > 1$.
(Its inverse is $(G^{i j}) = (\delta_{i j} - \frac{1}{k-1}$.)

Now, we prove that inequalities (\ref{r4.9a}), (\ref{r4.9b}) are obeyed
if   
 \begin{equation}
  x \neq  -  (m - 2) = x_c,
 \label{r4.12a} 
 \end{equation}
or $h \neq -  (m - 2) H$.

Let us suppose that  (\ref{r4.9a}) does not take place, i.e.  
$S_{HH} = - \frac{1}{2 \alpha}$.  Then using (\ref{r3.5}) we obtain  
\begin{equation}
S_{HH} - Q = - 2 (H - h) ((m-2) H +  h ) = 0, 
\label{r4.10a}                                            
\end{equation}
which implies  due to $H - h \neq 0$ (see  (\ref{r3.3}))         
\begin{equation}
   (m-2) H +  h = 0.
\label{r4.11a}                                                        
\end{equation}
 This relation contradicts to the  restriction (\ref{r4.12a}). 
 The obtained contradiction proves the inequality (\ref{r4.9a}).     
      
 Now let us  suppose that  (\ref{r4.9b}) is not valid,
 i.e. $S_{hh} = - \frac{1}{2 \alpha}$. Then using (\ref{r3.5}) we find  
 \begin{equation}
 S_{hh} - Q =  - 2 (h - H) (m -1) H  = 0. 
 \label{r4.10b}                                            
 \end{equation}
 But this relation contradicts  $H > 0$ and $H - h \neq 0$.
 This contradiction leads  us to the proof of the inequality
 (\ref{r4.9b}).         

Thus, we have proved that relations (\ref{r4.9a}) and (\ref{r4.9b}) are valid and hence the
 restriction (\ref{r4.2}) is satisfied for our solutions. 

Thus, we have proved the following proposition. 

{\bf Proposition 2.} {\em The cosmological solutions under consideration, 
which obey  $x = h/H \neq x_i$, $i = a,c,d$, where $x_a =1$, $x_c = - (m-2)$,  $x_d = - \frac{m}{2}$, 
are  stable if i) $x > x_d$ and unstable if ii) $x < x_d$. }


Now we  calculate  the number of non-special stable solutions  (i.e. obeying $x \neq x_c$) which are
given by Proposition 2 (see item i)). We denote this number as $n_{*}(\Lambda, \alpha)$.
By using the results  from the previous section (e.g. illustrated by figures) we 
obtain for $\alpha > 0$:   

 { \bf (1)  $m > 4$}          
  \begin{equation}
    n_{*}(\Lambda, \alpha) = \begin{cases}
    0, \  \Lambda \alpha \geq \lambda_d, \\
    1, \  \Lambda \alpha < \lambda_d; \\
    \end{cases}   \label{r4.13c}
  \end{equation}

  {\bf (2) $ m = 4$} 
     \begin{equation}
    n_{*}(\Lambda, \alpha) = \begin{cases}
    0, \  \Lambda \alpha \geq \lambda_d = \lambda_c, \\
    1, \   \Lambda \alpha < \lambda_d = \lambda_c ; \\
   \end{cases}   \label{r4.14c}
  \end{equation}

  {\bf (3) $ m = 3$} 
       \begin{equation}
      n_{*}(\Lambda, \alpha) = \begin{cases}
      0, \   \Lambda \alpha \geq \lambda_c, \\
      2, \   \lambda_d < \Lambda \alpha < \lambda_c, \\
      1, \   \Lambda \alpha \leq \lambda_d. \\
       \end{cases}   \label{r4.14cc}
    \end{equation}

 We see, that for $\alpha > 0$ and small enough value of $\Lambda$ there exists at least one stable solution 
 with $x \in (x_{d},x_{*})$. 
 
 {\bf Bounds on  $\Lambda \alpha$ for stable solutions with $\alpha > 0.$}
 Summarizing all cases presented above we find that for $\alpha > 0$
 stable exact solutions under consideration exist if and only if 
  \begin{equation}
      \Lambda \alpha < \begin{cases}
      \    \lambda_d  \ {\rm for} \ m \geq 4, \\
      \    \lambda_c  \  {\rm for} \ m = 3,
       \end{cases}   \label{r3.34ss}
    \end{equation}
 where $\lambda_c = \lambda_c (m)$ and  $\lambda_d = \lambda_d (m)$ 
 are defined in  (\ref{r3.22c}) and  (\ref{r3.22d}), respectively.

 In the case  $\alpha < 0$ we obtain       
    \begin{equation}
      n_{*}(\Lambda, \alpha) = \begin{cases}
      2, \  \Lambda |\alpha| > |\lambda_{a}|, \\
      0, \   \Lambda |\alpha| \leq |\lambda_{a}|, \\
     \end{cases}   \label{r4.17}
    \end{equation}
 i.e.  $ n_{*}(\Lambda, \alpha) =  n(\Lambda, \alpha)$. 
  Here the inequality  $x \neq x_a =1$ was used. Thus,   for $\alpha < 0$ and big enough value of $\Lambda$
 there exist two stable solution corresponding to  $x = x_1, x_2$ which obey $x_{*} < x_1 < x_2$. 
 
 {\bf Bounds on  $\Lambda |\alpha|$ for stable solutions with $\alpha < 0.$}
 It follows from (\ref{r4.17}) that for $\alpha < 0$
 stable exact solutions  under consideration exist if and only if
 the relation (\ref{r3.36s})  is obeyed.

\section{Solutions describing a small enough variation of $G$}

  Here we analyze the solutions  by using the restriction on variation of the effective
  gravitational constant $G$, which is inversely proportional (in the Jordan frame) 
  to the  volume scale factor  of the (anisotropic) internal space \cite{IvKob}, i.e.

  \begin{equation}
  \label{r5.G0}
    G = {\rm const } \exp{[- (m - 3) H t - 2 h t]}.
   \end{equation}
 
By using (\ref{r5.G0}) one can get the following formula 
for a dimensionless parameter of temporal variation of $G$ ($G$-dot):
\begin{equation}
  \delta \equiv \frac{\dot{G}}{GH}  =  - (m - 3 + 2 x), \qquad x = h/H.
\label{r5.G}
\end{equation}
Here  $H >0$ is the Hubble parameter.

Due to observational data, the variation of the gravitational constant is 
on the level of $10^{-13}$ per year and less.

For example, one can use, as it was done in ref. \cite{IvKob},  the following bounds on the 
value of the dimensionless variation of the  effective gravitational constant:

 \begin{equation}
 \label{r5.G1}
  - 0,65 \cdot 10^{-3} < \delta < 1,12 \cdot 10^{-3}.
 \end{equation}
They come from the most stringent limitation
on $G$-dot obtained by the set of ephemerides \cite{Pitjeva}
and  value of the Hubble parameter (at present) \cite{Ade}
 when both are written with 95\% confidence level \cite{IvKob}.

  When the value  $\delta$ is fixed we get from (\ref{r5.G})
    \begin{equation}
  x =  x_0(\delta,m) \equiv  - \frac{(m - 3 + \delta)}{2}. 
  \label{r5.0}    
 \end{equation}

We obtain 
   \begin{equation}
   x_0(\delta,m) =   - (m - 3 + \delta)/2 > x_{*}(m) =  - (m - 2)/2  
   \label{r5.00}    
  \end{equation}
for $ \delta < 1$,  e.g. when restrictions (\ref{r5.G1}) are obeyed.
In this case 
\begin{equation}
   {\cal P}(x_0(\delta,m),m) = (m - 1)(1 - \delta)  > 0
   \label{r5.P}    
\end{equation}

and hence (see (\ref{r3.8}) )
 \begin{equation}
   \alpha < 0.
   \label{r5.alpha}    
  \end{equation}

Due to  inequality (\ref{r5.00}) and relations $x_c(m) < x_{*}(m)$,  $x_d(m) < x_{*}(m)$ 
all conditions in Proposition 2 are satisfied  for $ \delta < 1$, e.g.  when 
restrictions  (\ref{r5.G1}) are obeyed. This implies the stability of the
solutions under consideration ($m > 2$) with small enough variation of $G$, 
which obey  the physical bounds (\ref{r5.G1}).   
 
We note that for $ \delta = 0$, i.e. for solutions with zero variation of
$G$, we obtain  from (\ref{r3.8L}) 
\begin{equation}
    \Lambda |\alpha| = - \lambda (x_0(0,m))  =  \frac{2m^2-5m+9}{8(m-1)} > 0, 
   \label{r5.Lambda}    
 \end{equation}
$m > 2$. For $m =3,4,5$ we find 
  \begin{equation}
      \Lambda |\alpha| = \begin{cases}
      \    \frac{3}{4} \ {\rm for} \ m =3, \\
      \    \frac{7}{8}  \ {\rm for} \ m = 4, \\
      \   \frac{17}{16}  \ {\rm for} \ m = 5
       \end{cases}   \label{r5.L345}
 \end{equation}
  with   
  \begin{equation}
          x_0(0,m) = - (m - 3)/2  =  \begin{cases}
          \    0 \ {\rm for} \ \ m =3, \\
          \    - \frac{1}{2}  \ {\rm for} \ m = 4, \\
          \    -1, \ {\rm for} \ m = 5.
           \end{cases}   \label{r5.l345}
  \end{equation}

\section{Conclusions}

We have considered the  $D$-dimensional  Einstein-Gauss-Bonnet (EGB) model
with the $\Lambda$-term.  
We have found a new class of cosmological   solutions with diagonal  metrics and exponential time dependence of two scale factors. The solutions are governed by two Hubble-like parameters $H >0$ and $h$, 
corresponding to flat subspaces of dimensions $m > 2$ and $2$, respectively ($D = m + 3$). 
Here  the  parameters $H$ and $h$ satisfy the 
following restrictions:  $H \neq h$ and  $ mH + 2h \neq 0 $. The solutions obey the relation 
$ (m-2)H + 2h \neq 0 $.

Any obtained solution may be considered as describing an exponential expansion of  $3$-dimensional subspace  (``our space'') with  the Hubble parameter $H > 0$ and anisotropic behaviour of $((m -3) + 2)$-dimensional anisotropic internal space which expands in $(m-3)$ dimensions (with Hubble-like parameter $H$) and  either contracts, 
or expands (with Hubble-like parameter $h$) or stable in $2$ dimensions.
The solutions are governed by a polynomial equation of  third order  in $x = h/H$, which 
is readily solved in radicals for all values of $\Lambda$ and $\alpha$ in Appendix A. Here $\alpha = \alpha_2 / \alpha_1 $ where $\alpha_1$ and $\alpha_2$ are two non-zero constants of the model.  
The case $\Lambda = 0$, which takes place only for $\alpha > 0$, is outlined in Appendix B.   We note that in our previous paper \cite{IvKob-GRG-18}, devoted to $(1+ m + l)$-dimensional solutions with $m > 2$ and $l > 2$, the polynomial master equation in $x = h/H$ was generically of fourth order.
Here we have found the restrictions on  $\Lambda$ which guarantee the existence
of the  solutions under consideration:
 \begin{equation}
    \Lambda \alpha \neq 1/4,  \quad {\rm if } \  m = 4
      \label{c.1}
   \end{equation}
 for $\alpha > 0$ and
 \begin{equation}
     \Lambda |\alpha| >    \frac{(D -2)(D-1)}{8(D - 4)(D - 3)} 
              \label{c.2}
   \end{equation}
for $\alpha < 0$.  
  
Using the scheme  from  ref. \cite{Ivas-16},  
we  have proved that any of these solutions obeying the additional restriction 
$x \neq  x_c = - (m -2)$,
 is stable (as  $t \to + \infty$) if  $x > x_d = - m/2$ and unstable if  $x < x_d$.

We have also found that for $\alpha > 0$  stable exact solutions
exist if and only if: 
\begin{equation}
 \Lambda \alpha < \lambda_d  \label{c.3}
 \end{equation}
 for $m \geq 4$    and  
 \begin{equation}
 \Lambda \alpha <  \lambda_c,  \label{c.4}
 \end{equation}
 for $m < 4$,     
 where   $\lambda_d$ is defined in   (\ref{r3.22d}) and $\lambda_c$ is defined in   (\ref{r3.22c}).
 For $\alpha < 0$ the all obtained exact solutions (obeying (\ref{c.2})) are stable.

We have  shown that all (well-defined) solutions with small enough varation of 
the effective gravitational constant $G$ (in the Jordan frame) are stable.

The solutions   presented  here are different from  those obtained in Ref. \cite{IvKob-GRG-18}  for two non-coinciding Hubble-like parameters $H > 0$ and $h$ corresponding to factor spaces of dimensions $m > 2$ and $l > 2$ with $mH +  l h \neq 0$.   
In the case of  Ref. \cite{IvKob-GRG-18} we have two branches with: (a) $\alpha > 0$, $ - \infty <  \Lambda \alpha < \lambda_{+}(m,l)$ and  (b) $\alpha < 0$, $  \Lambda |\alpha| > \lambda_{-}(m,l)$, where $ \lambda_{\pm}(m,l) > 0$.  
In the case $l= 2$ and $\alpha > 0$ the allowed values for $\Lambda$ are unbounded from the top and the bottom but they are bounded from the top when  stable solutions are considered. For $l=2$,  we have   $x = h/H < x_{*} = -(m-2)/2$ for $\alpha > 0$  and $x  > x_{*}$ for $\alpha > 0$,  while in the case $l>2$ we get: $x_{-}  < x < x_{+} < 0$ for $\alpha > 0$,  and $x < x_{-} $ (unstable brunch)  or $ x > x_{+}$ (stable brunch) for $\alpha < 0$.


\renewcommand{\theequation}{\Alph{subsection}.\arabic{equation}}
\renewcommand{\thesection}{}
\renewcommand{\thesubsection}{\Alph{subsection}}
\setcounter{section}{0}

\section{Appendix}

\subsection{The analytical solution to the master equation}

For any $m  > 2$ the master equation (\ref{r3.13.M})  may 
be written as follows
 \begin{equation}
   x^3 +  b x^2 +  c x + d = 0,    \label{A.1}  
               \qquad    
 \end{equation} 
where
\begin{eqnarray} 
b = m - 1 + 4 \lambda (m - 1), \qquad \qquad    \\ 
c = (1/2)(m-1)m + 4 \lambda(m - 1)(m - 2), \label{A.1c} \\
d = (1/8)(m-2)m(m+1) + \lambda (m - 1)(m - 2)^2.  \label{A.1d}
\end{eqnarray} 

The standard (generically complex) Cardano solution to equation (\ref{A.1}) gives us 
 \begin{equation}
 x = x_k = - b/3 + e_k \sqrt[3]{Z} - R/(e_{k} \sqrt[3]{Z}),  \label{A.2x} 
  \end{equation} 
 where
 \begin{eqnarray}
 Z = (1/2) 3^{-3/2} (\pm \sqrt{W}) + Q, \label{A.2Z} \\
 W=  27 d^2+(4b^2-18c)bd + 4c^3 - b^2 c^2, \label{A.2W} \\
 R = c/3 - b^2/9, \qquad  \qquad  \label{A.2R} \\
 Q = (bc-3d)/6 - b^3/27, \label{A.2Q }  
\end{eqnarray}
 and $e_k =  \exp(2 \pi i k/3)$ are cubic roots from unity, $k = 0,1,2$. 
 Here for a given value of square root $\sqrt{W}$ one should choose the sign $\pm$
such that $Z \neq 0$ and   $\sqrt[3]{Z}$ is arbitrary (fixed) value of cubic root. 

The calculations give us
\begin{eqnarray}
W= ((32q-m^2-2m)(8mq+m^2+3m+2) \times \nonumber\\
 \times (8(m -2)q -3m^2+13m-16))/64, \label{A.2WW} \\ 
R = (-32q^2 + 8(m - 4)q+ (m+2)(m-1))/18 \label{A.2RR},\\
Q = (-1024q^3+384(m-4)q^2 +24(m^2+10m-20)q  \nonumber \\
    -(m+2)(7m^2-17m-8))/432, \label{A.2QQ}  
\end{eqnarray}
where $q = \lambda (m-1) = \Lambda \alpha (m-1)$.

Here $W = - \Delta$, where $\Delta$ is discriminant corresponding to (\ref{A.1}).  
We get   
\begin{equation}
 W= 32 (m-1)^3 m (m -2) (\lambda - \lambda_a) (\lambda - \lambda_c) (\lambda - \lambda_d), 
  \label{A.3W}
\end{equation}
where $\lambda_a,  \lambda_c, \lambda_d$  are defined in (\ref{r3.22a}), (\ref{r3.22c}),  (\ref{r3.22d}),
respectively.

It is known that for $\Delta > 0$ the equation (\ref{A.1}) has three different real solutions,
while for $\Delta < 0$ it has is only one real solutions. For $\Delta = 0$ we have: either 
i) two different real solutions (two roots are coinciding) or ii) or one real solutions 
(all three roots are coinciding). One can readily verify that these facts  confirm  the classifiction 
of the number of solutions presented in Section 3 (up to exclusion one of the solutions for 
$\lambda = \lambda_d$ corresponding to $x = x_d$.) We note that the case  
$\Delta = 0$ takes place when $\lambda = \lambda_i$ for some $i = a,c,d$. 
In the subcase ii) we obtain solutions: $x = x_a =1$ (for $\alpha < 0$, $\lambda = \lambda_a$) and 
$x = x_d = x_c = -2$ (for $\alpha > 0$, $m = 4$ and $\lambda = \lambda_c = \lambda_d = 1/4$), 
which are forbidden by our restrictions.

\subsection{The case $\Lambda  = 0$}

 For $\Lambda \alpha = \lambda = 0$ we obtain  
\begin{eqnarray}
 W=  m(m+1)(m+2)^2(3m^2 -13m +16)/64 = \nonumber \\
  32 (m-1)^3 m (m -2) (- \lambda_a) \lambda_c  \lambda_d  >0   \label{A.4W}
\end{eqnarray}
for $m > 2$ or $\Delta < 0$. Hence in this case we have only one real solution to master equation  
which is agreement with our graphical analysis from Section 3. This solution takes place only for
$\alpha >0$. It is given by the relations (following from those presented above)
\begin{eqnarray}
x = \sqrt[3]{Z} - R/\sqrt[3]{Z} - (m-1)/3, \qquad  \label{A.5x} \\
R = (m-1)(m+2)/18,  \qquad  \qquad  \qquad           \label{A.5R} \\
Z = 2^{-4} 3^{-3/2} (m+2)\sqrt{m(m+1)(3m^2-13m+16)}  + Q, \label{A.5Z} \\
Q = -((m+2)(7m^2-17m-8))/432, \qquad \label{A.5Q}
\end{eqnarray}
$m > 2$. It may be shown that $Z >0$ for all natural $m$.
The calculations give us the following approximate values:
\begin{equation}
      x \approx \begin{cases}
        - 0.72212, \  {\rm for} \ m = 3, \\
        - 1.32219   \  {\rm for} \ m = 4,  \\
        - 1.86971   \  {\rm for} \ m = 5. 
     \end{cases}   
    \end{equation}

\vspace{0.2truecm}

 {\bf Acknowledgments}

This paper has been supported by the RUDN University Strategic Academic Leadership Program (recipient V.D.I., mathematical model development).  The reported study was funded by RFBR, project number 19-02-00346 
(recipient A.A.K., simulation model development).


\end{document}